\documentclass[
  reprint,
  amsmath,amssymb,
  aps
]{revtex4-2}

\usepackage[T1]{fontenc}
\usepackage{graphicx}
\usepackage{dcolumn}
\usepackage{bm}
\usepackage{dsfont}
\usepackage{placeins}      
\usepackage{dblfloatfix}   
\usepackage{microtype}
\usepackage{physics}

\makeatletter
\newcommand{\paneltagH}[2]{%
  \makebox[0pt][l]{\raisebox{#1}[0pt][0pt]{\hspace{0.01\textwidth}\textbf{(#2)}}}%
}
\makeatother

\begin{document}

\preprint{APS/123-QED}

\title{Flux-Switching Floquet Engineering}

\author{Ian Emmanuel Powell}
\author{Louis Buchalter}
\affiliation{Physics Department, California Polytechnic State University San Luis Obispo California.}

\date{\today}

\begin{abstract}
We present an analysis of a square-lattice Harper-Hofstadter model with a periodically varying magnetic flux with time. By switching the dimensionless flux per plaquette between a set of values $\{p_j/q_j\}$ the Floquet quasienergy spectrum is folded into $Q$ = lcm$\{q_j\}$ bands.  We determine closed-form analytical solutions for the quasienergy spectrum and Chern numbers for the $-1/2 \rightarrow 1/2$ flux switching case, as well as the Rudner-Lindner-Berg-Levin (RLBL) winding invariants $W$ numerically, and construct the corresponding topological phase diagram for arbitrary driving period.  We find that generic flux-switching drives feature interlaced Hofstadter butterfly quasienergy spectra which may host Floquet topological phases with no static counterpart, and the gaps in the spectrum may be labeled according to a Diophantine equation which relates the quasienergy gap index to the fluxes attained in the drive and their associated per-step windings.

\end{abstract}

\maketitle

\section{Introduction\label{sec:intro}}

Periodically driven, or Floquet, quantum matter provides a
powerful arena in which band structures, symmetries and even
dimensionality can be tuned on demand
\cite{Goldman2014, Oka2019, Eckardt2017}. These periodically driven systems can display a wide variety of rich phenomena such as dynamical localization \cite{Dunlap1986}, coherent destruction of tunneling \cite{Grossmann1991}, and the emergence of anomalous topological phases whose bulk Chern numbers vanish even as robust chiral edge modes persist \cite{Kitagawa2010, Rudner2013, Nathan2019}.
 Experimental protocols have successfully implemented periodic driving to realize the novel non-equilibrium physics involved: photonic waveguide arrays with helically modulated couplings have been used to realize topological phases \cite{Rechtsman2013}, and ultracold atoms in shaken optical lattices have been used to realize the Floquet Haldane model \cite{Jotzu2014_FHaldane}. More recently, signatures of the so-called Floquet anomalous topological insulators have been inferred via edge transport imaging \cite{Maczewsky2017}, the light-induced anomalous Hall effect has been directly measured in graphene \cite{McIver2020}, and higher-order Floquet topological states have been realized in three-dimensional acoustic lattices \cite{Zhu_corner}.

In this manuscript, we investigate a periodically driven Harper–Hofstadter model in which the dimensionless flux per plaquette is toggled between distinct rational values over each driving period. This time-periodic flux modulation fragments the quasienergy spectrum into a structured set of magnetic subbands, leading to a vast nontrivial topological phase diagram in which quasienergy gaps are labeled by their Rudner-Lindner-Berg-Levin (RLBL) winding invariants $W$, including ``anomalous'' windings in which chiral edge modes propagate through the Floquet zone edge. We further show that the global organization of gaps follows a compact Diophantine congruence that links each gap label to step-resolved winding contributions. Although our analysis primarily focuses on noninteracting charged fermions on a square lattice, the same Floquet framework applies directly to bosonic cold-atom setups: by modulating Raman-laser detunings and intensities in time, one can realize our flux-switching protocol on existing experimental platforms \cite{Aidelsburger2013flux,Miyake2013flux}.  

\section{\label{sec:theory}Theory}
We consider a model of free electrons on a square lattice that are subject to a periodic flux switching routine in which the dimensionless flux per plaquette switches between rational $\alpha = p/q$ at regular intervals in time.  For example, for two-flux routines, the electrons are subject to two distinct magnetic fields that are perpendicular to the plane of the lattice--$\bf{B_1}$, with associated dimensionless flux per plaquette $\alpha_1$, for a time $T_1$, followed by a magnetic field $\bf{B_2}$, with associated dimensionless flux per plaquette $\alpha_2$, for a time $T_2$, and so on.  The time evolution operator for one full period of the drive is given in natural units as
\begin{equation}
\label{eq: 1}
U(T+t_0,t_0)
=\mathcal{T}\exp\Bigl[-i\!\int_{t_0}^{t_0+T}H(t)\,\mathrm{d}t\Bigr],
\end{equation}
where $T$ is the period of the drive, and \(\mathcal{T}\) denotes time‐ordering.  Given a sequence of flux quenches with $L$ distinct steps,  i.e. for the case where the magnetic flux discontinuously jumps between different values $L$ times, and taking the reference time $t_0 = 0$, eq. \eqref{eq: 1} factorizes to the time-ordered product of constituent unitaries
\begin{equation}
\label{eq:U_piecewise}
U(T)
=\prod_{j=1}^L\exp\bigl[-i\,H_j\,T_j\bigr]
=\exp\bigl[-i\,H_F\,T\bigr],
\end{equation}
where $T = \sum_j T_j$, $H_j$ is the Hamiltonian that describes the electrons during the j$^{\text{th}}$ step of the drive, and \(H_F\) is the effective Floquet Hamiltonian--the generator of the stroboscopic dynamics.  Given the definition of $H_F$ featured in Eq. \eqref{eq:U_piecewise} the eigenvalues, or ``quasienergies,'' of $H_F$ are defined modulo \(2\pi/T\).  

The constituent Hamiltonians, $H_j$, in Eq. \eqref{eq:U_piecewise} describe non-interacting electrons in the presence of an external magnetic field, or non-interacting bosons in the presence of a synthetic gauge field arising from laser induced Raman-assisted tunneling.  They read 
\begin{equation}
H_j = -\sum_{ \mathbf{r}, \mathbf{r}'} t_{\mathbf{r}, \mathbf{r}'}\,e^{-i \phi_{j, \mathbf{r}, \mathbf{r}'} }c^\dagger_\mathbf{r} c_{\mathbf{r}'} +\text{h.c.},
\end{equation}
where $t_{\mathbf{r}, \mathbf{r}'}$ is the hopping integral from site $\mathbf{r}$ to $\mathbf{r}'$, $c^\dagger_\mathbf{r}$ and $c_{\mathbf{r}'}$ are the electron (boson) particle creation and annihilation operators at the sites $\mathbf{r}$ and $\mathbf{r}'$ respectively, spin indices have been suppressed, and $\phi_{j, \mathbf{r} , \mathbf{r}'}$ is the Peierls phase acquired by the hopping of particles in the presence of the magnetic field (synthetic gauge field) 
$\phi_{j,\mathbf{r'},\mathbf{r}} = \tfrac{e}{\hbar}\int_\mathbf{r}^{\mathbf{r}'} \mathbf{A}_j(\mathbf{l}) \cdot d\mathbf{l}  $.  Upon selecting the Landau gauge $\bf{A_j}$$ = (-B_j y, 0, 0)$ and including nearest and next-nearest neighbor hopping terms, the Hamiltonians are explicitly written as
\begin{equation}
\label{eq: H_j}
H_{j}=H_{j}^{\text{NN}}+H_{j}^{\text{NNN}},
\end{equation}
with
\begin{align}
H_{j}^{\text{NN}}
  &= -t\,\!
     \sum_{m,n}
     \Bigl[
        e^{-\mathrm i 2\pi\alpha_{j} n}\;
        c_{m+1,n}^{\dagger}c_{m,n} +c_{m,n+1}^{\dagger}c_{m,n}
     \Bigr]
     +\text{h.c.},
     \label{eq:H_nn} 
\end{align}
\begin{align}
H_{j}^{\text{NNN}}
  &= -t'\,\!
     \sum_{m,n}
     \Bigl[
        e^{-\mathrm i2\pi\alpha_{j}(n+1/2)}
        c_{m+1,n+1}^{\dagger}c_{m,n} \notag\\
  &\hspace{3.8em}
      +\,e^{-\mathrm i2\pi\alpha_{j}(n-1/2)}
        c_{m+1,n-1}^{\dagger}c_{m,n}
     \Bigr]
     +\text{h.c.},
     \label{eq:H_nnn}
\end{align}
where \(\alpha_{j}=p_{j}/q_{j}\) (with coprime $p_j$ and $q_j$) is the reduced dimensionless flux per plaquette in step~\(j\), and
the lattice coordinates on the square lattice are given via \(\mathbf r = m a\,\hat{\mathbf x}+n a\,\hat{\mathbf y}\) with lattice constant $a$.

For each step in the time evolution of the system subject to a sequence of fluxes $\{\alpha_j\}=\{p_j/q_j\}$ the lattice magnetic translations
$\hat{T}_x,\hat{T}_y$ satisfy Zak’s algebra \cite{brown1964,zak1964a,zak1964b}
\begin{equation}
\hat{T}_x \hat{T}_y \;=\; e^{i2\pi\alpha_j}\, \hat{T}_y \hat{T}_x,
\qquad \alpha_j=\frac{p_j}{q_j}.
\label{eq:zak-algebra}
\end{equation}
Hence $[\hat{T}_x^\ell,\hat{T}_y^h]=0$ if $e^{i2\pi \alpha_j \ell h}=1$.  Consequently, the minimal area supercell compatible with all steps in the drive must have area \( \ell h \) (in units of $a^2$) that is an integer multiple of the least common multiple of the set of all $\{q_i\}$--i.e the minimal cell has area \( Q \equiv \operatorname{lcm}\{q_j\} \).

Selecting $(\ell,h)=(1,Q)$, the magnetic translations $\hat{T}_x$ and $\hat{T}_y^Q$ commute with each $H_j$ (and thus with each $U_j$). This choice defines a magnetic unit cell that is $Q$ times longer along $y$.
With this choice one has, for every step $j$,
\begin{equation}
\hat{T}_y^{\,Q} U_j (\hat{T}_y^{\,Q})^\dagger = U_j, \qquad [U_j,\,\hat{T}_x]=0.
\end{equation}
Inserting $\hat{T}_y^{\,Q}(\hat{T}_y^{\,Q})^\dagger$ between steps in the Floquet product straightforwardly
yields
\begin{equation}
[\hat{T}_y^{\,Q},\,U(T)]=0, \qquad [\hat{T}_x,\,U(T)]=0.
\end{equation}
We therefore simultaneously diagonalize $U(T)$, $\hat{T}_x$, and $\hat{T}_y^{\,Q}$. The
quasienergy spectrum splits into $Q$ magnetic subbands, and Bloch’s theorem
holds in the reduced (magnetic) Brillouin zone (RBZ)
\begin{equation}
k_x\in\left(-\frac{\pi}{a},\,\frac{\pi}{a}\right], \qquad
k_y\in\left(-\frac{\pi}{Qa},\,\frac{\pi}{Qa}\right].
\end{equation}

Before we proceed, a couple of remarks regarding time-varying flux are in order.  First, it should be noted that introducing finite, continuous ramps of duration $\tau$ between the piecewise steps in Eq.~\eqref{eq:U_piecewise} formally breaks the exact magnetic translation symmetry of the ideal case of a series of quenches. However, provided $\tau/T <<  1$ these ramps do not close the primary quasienergy gaps, so all associated topological features remain intact. Consequently, although we omit the ramping terms from our main analysis, our results remain valid in the appropriate small‐ramp‐time limit where primary gaps are not destroyed. We demonstrate the effect of including ramping terms for a model driven between -1/2 and 1/2 flux in Appendix \ref{app: ramps}. Second, by Faraday’s law, any temporal change in flux induces an electric field in electronic systems.  In cold‐atom Raman‐assisted‐tunneling implementations the time varying gauge field manifests as an analogous synthetic electric field \cite{Lin2011}.  In either realization identical stroboscopic Floquet-Bloch Hamiltonian matrices are produced.  

In terms of experimental realizability, Moiré heterostructures (e.g., twisted bilayer graphene) may host a large flux per unit cell, but to realize access the high-frequency (and intermediate frequency) regimes, one must alternate the magnetic flux at ultrafast scales set by the band gap widths. With \(t\!\sim\!1~\mathrm{meV}\) and our dimensionless choice for the high frequency cutoff \(T\!\sim\!1\), the drive frequency is \(\omega\!\sim\!t/\hbar \approx 1.5\times 10^{12}~\mathrm{s^{-1}}\).  By contrast, Raman-assisted tunneling in cold atom systems can realize uniform synthetic flux with \(\alpha\) continuously tunable over \([0,1)\) ~\cite{Miyake2013flux,Aidelsburger2013flux}, and for typical parameters $t/\hbar \sim .1 - 1$ kHz, kHz drive frequencies correspond to $T \sim \mathcal{O}$(1) in our dimensionless units--thus the high and intermediate frequency regimes are much more accessible.  Furthermore, higher frequency driving leads to exponentially long prethermal plateaus \cite{huveneers}, making direct measurements more feasible.  These considerations make cold-atom platforms the most natural testbed for our flux-switching protocol.

\subsection{Topology and Chiral Edge States}
For this 2 + 1 dimensional time periodic system the Floquet-Bloch time evolution operator, $U(k_x, k_y, t)$, defines a map from the three-torus (RBZ$\times S^1$) to the unitary group of $Q\times Q$ matrices, $U(Q)$, where again $Q = \operatorname{lcm}\{q_j\}$.  This map carries a winding known as the RLBL winding, $W_\varepsilon$, which accurately predicts the number of chiral edge modes propagating through each gap at $\varepsilon$ in the quasienergy spectrum \cite{Rudner2013}.  The first Chern number of the occupied bundle, $C$, which is featured in the familiar quantization of the integer quantum Hall conductivity \cite{TKNN}, is the total Berry flux through the RBZ; it accurately predicts the number of chiral edge modes propagating through each gap provided no ``anomalous'' modes cross the Floquet boundary typically taken to be $\varepsilon = \pm \tfrac{\pi}{T}$ in the principal zone.  As a rule of thumb,  when each instantaneous Hamiltonian's operator norm satisfies
$\|H_j\|\,T \;<\;\pi$,
no anomalous winding occurs and the Chern numbers of the quasienergy bands suffice to fully characterize the topology. On the other hand, When $\|H_j\|T\ge\pi$, anomalous modes at the Floquet zone boundary may appear, and one must compute the RLBL winding, $W_\varepsilon$, to obtain the correct edge-mode count.

To define $W_\varepsilon$ we proceed in the usual way by defining the bulk time evolution operator
\(U(\mathbf k,t)=\mathcal T\exp\!\left[-i\!\int_{0}^{t} H(\mathbf k,t')\,dt'\right]\),
and the branch cut-specified 
Floquet Hamiltonian by
\begin{equation}
U(\mathbf k,T)=e^{-iT\,H_F^{(\varepsilon)}(\mathbf k)},\qquad
H_F^{(\varepsilon)}(\mathbf k)=\frac{i}{T}\,\log_{-\varepsilon T}U(\mathbf k,T),
\label{eq: branchcut_ham}
\end{equation}
where log$_{-\varepsilon T}$ denotes the logarithm with the branch cut located at $e^{-i\varepsilon T}$--i.e. log$_{-\varepsilon T}(e^{i\phi}) = i\phi$ for $-\varepsilon T - 2\pi < \phi \leq -\varepsilon T$.
The periodized evolution is then defined as the unitary
\begin{equation}
V_\varepsilon(\mathbf k,t)=U(\mathbf k,t)\,e^{+it\,H_F^{(\varepsilon)}(\mathbf k)},
\end{equation}
which is smooth on the three-torus $\mathbb T^3=\mathrm{RBZ}\times S^1$ and satisfies
$V_\varepsilon(\mathbf k,0)=V_\varepsilon(\mathbf k,T)=\openone$.
The RLBL winding of the gap at \(\varepsilon\) is \cite{Carpentier2015}
\begin{equation}
W_\varepsilon
= \frac{1}{24\pi^2}\int_{\mathrm{RBZ}\times S^1}
\mathrm{Tr}\,\bigl[\!(V_\varepsilon^{-1} dV_\varepsilon)^3\bigr],
\label{eq:RudnerW}
\end{equation}
which equals the net number of chiral edge modes crossing the quasienergy gap at $\varepsilon$. If \(\{\varepsilon_n\}\) lists the gaps in order (including that at the bottom boundary of the principal Floquet zone, \(-\pi/T\)),
the band Chern numbers \(\{C_n\}\) satisfy the telescoping relations \cite{Rudner2013}
\begin{equation}
C_n=W_{\varepsilon_{n+1}}-W_{\varepsilon_n},\qquad 
\sum_n C_n=0,
\label{eq:WminusWisC}
\end{equation}

To evaluate $W_\varepsilon$ one can either count the number of chiral edge modes propagating through each gap, implement the Floquet-St\v{r}eda formula \cite{Streda}, or one may numerically evaluate the bulk winding of the mapping to $U$ via a cubic grid of $k_x, k_y, t$ values, $\Lambda$.  We implement all three methods in this manuscript; but in the following section we elect to numerically compute $W_\varepsilon$.  To this end, we implement the formula of Morikawa and Suzuki \cite{MorikawaSuzuki2024W3} which we briefly review here. The derivatives in Eq. \eqref{eq:RudnerW} are replaced via
\begin{equation}
V^{-1}_\varepsilon \partial_\mu V_\varepsilon \rightarrow \frac{1}{4}\Omega_\varepsilon(\mathbf{x},\mu),
\end{equation}
where $\mathbf{x}\ = (k_x,k_y,t) \in \Lambda$ and
\begin{equation}
\begin{aligned}
\Omega_\varepsilon(\mathbf{x},\mathbf{\mu})
&=V^{-1}_{\varepsilon}(\mathbf{x})\Bigl[
   V_{\varepsilon}(\mathbf{x}+\hat\mu)
 - V_{\varepsilon}(\mathbf{x}-\hat\mu)\\
&\quad
 -\tfrac{1}{6}\bigl\{
    V_{\varepsilon}(\mathbf{x}+2\hat\mu)
  -2\,V_{\varepsilon}(\mathbf{x}+\hat\mu)\\
&\quad\quad
  +2\,V_{\varepsilon}(\mathbf{x}-\hat\mu)
  -V_{\varepsilon}(\mathbf{x}-2\hat\mu)
   \bigr\}
  \Bigr]
  -\mathrm{H.c.}
  \label{eq: Suzuki}
\end{aligned}
\end{equation}
with $\mu = k_x, k_y, t$.  The RLBL winding (Eq. \eqref{eq:RudnerW}), is calculated on the discrete grid as approximately
\begin{equation}
\begin{aligned}
W_\varepsilon 
&\approx \frac{1}{4^3 24\,\pi^2}
   \sum_{\mathbf{x}\in \Lambda}
   \sum_{\mu,\nu,\rho}
   \epsilon_{\mu\nu\rho}\\
&\quad\times 
   \mathrm{Tr}\bigl[
     \Omega_\varepsilon(\mathbf{x},\mu)\,
     \Omega_\varepsilon(\mathbf{x},\nu)\,
     \Omega_\varepsilon(\mathbf{x},\rho)
   \bigr]\,.
   \label{eq:Suzuki_W}
\end{aligned}
\end{equation}
All calculations of $W_\varepsilon$ are performed on a $91\times91\times51$ grid when using the above expression and a demonstration of numerical convergence is done in Appendix \ref{app:convergence}.

\subsection{\label{sec:pi-pi} $(-1/2\rightarrow 1/2)$ Flux Switching}

\begin{figure}
  \paneltagH{0.225\textheight}{a}%
  \includegraphics[width=0.45\textwidth,height=0.22\textheight,keepaspectratio]{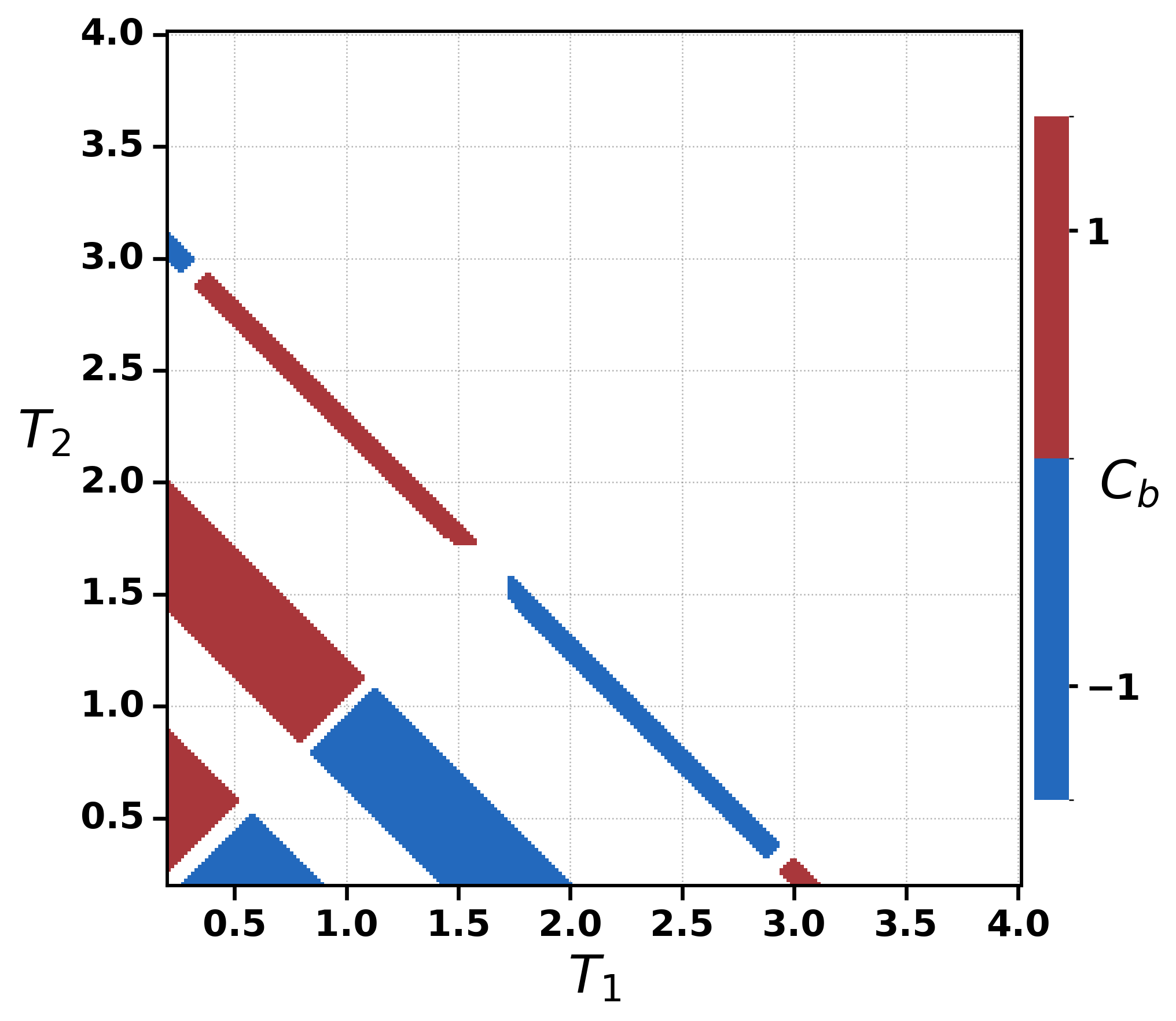}%
  \hfill
  \paneltagH{0.18\textheight}{b}%
  \includegraphics[width=0.45\textwidth,height=0.28\textheight,keepaspectratio]{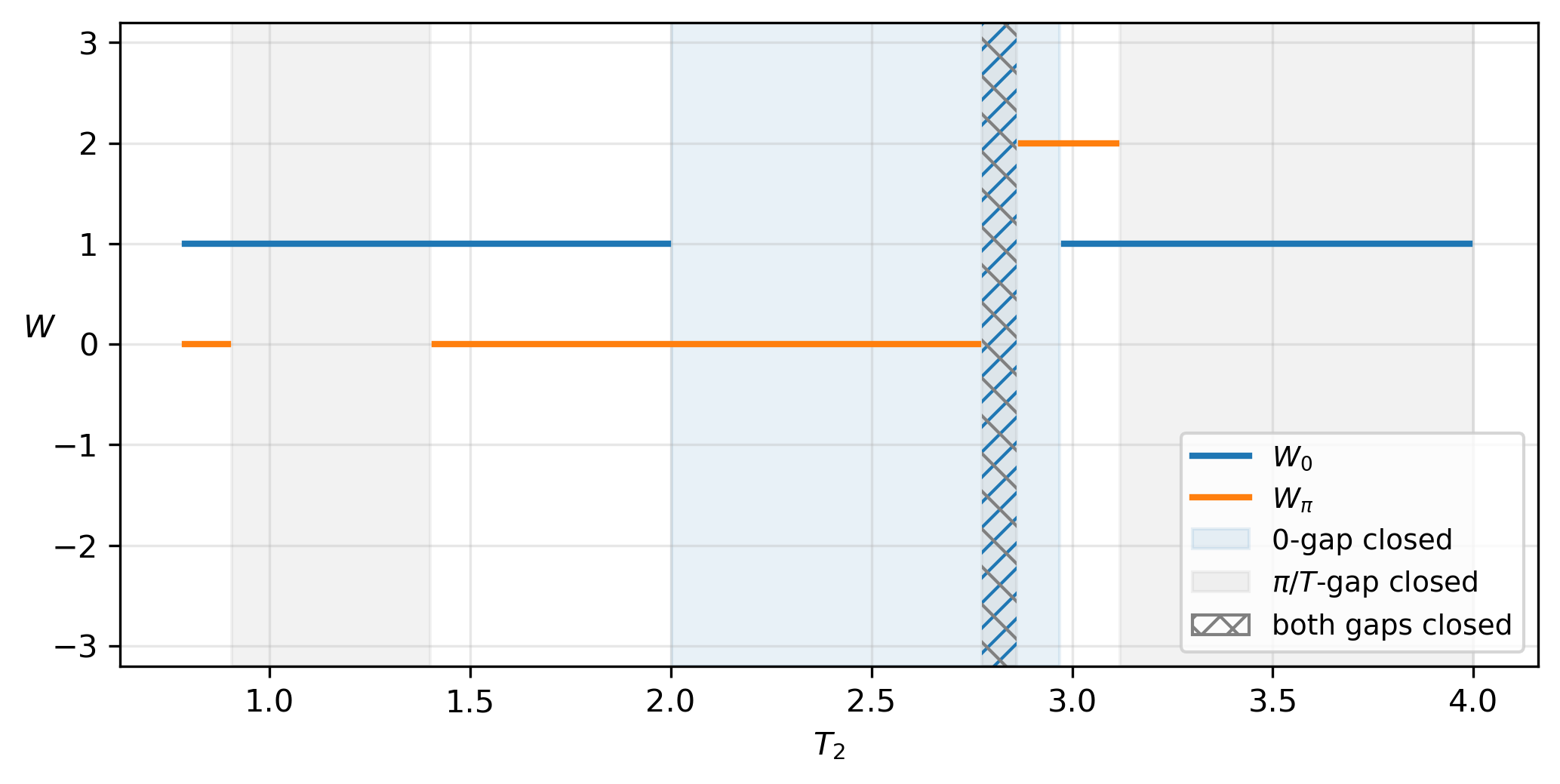}

  \caption{\raggedright
    (a) Chern number of the bottom band in the principal Floquet zone, $C_b$, as predicted by Eq.~\eqref{eq:Chern_sign_rule} plotted versus dwell times $T_1$ and $T_2$ with $t' = 0.3 t$. White areas are regions where the two bands are not well separated, and the origin corresponds to $T_1 = T_2 = 0.2$. Topological phases with nonzero $W_\pi$ occur for small windows of $T_1$, $T_2$ values: $T_1 \sim .2$, $T_2 \sim 3$, and $T_1 \sim 3$, $T_2 \sim .2$. (b) $W_{\pi}$ and $W_{0}$ calculated using Eq.~\eqref{eq:Suzuki_W} plotted as a function of $T_2$ with $T_1 = 0.2$; values deviate from $W = 0, 1, 2$ by a maximum of approximately $\sim \pm 10^{-2}$. A demonstration of the numerical convergence of Eq.~\eqref{eq:Suzuki_W} is featured in Appendix \ref{app:convergence}.}
  \label{fig:Chern_pi_pi}
\end{figure}

We begin by considering the simplest possible nontrivial two–step protocol where the flux is switched between $\alpha = -1/2$ and $\alpha = +1/2$ with a nonzero NNN hopping integral.  This drive admits a closed-form solution to the quasienergies in momentum space as well as the quasienergy band Chern numbers provided they are well separated from one another. Denoting the dwell times in the -1/2 and 1/2 flux steps by \(T_1\) and \(T_2\), the single-particle Bloch Hamiltonians in the $y$-even-odd sublattice basis are obtained via the appropriate Fourier transformations of Eq. \eqref{eq: H_j} as
\begin{equation}
\begin{split}
H_{\pm1/2}(\mathbf k)
&=-2t\,\mathbf h_{\pm1/2}(\mathbf k)\cdot\boldsymbol\sigma,\\
\mathbf h_{\pm1/2}(\mathbf k)
&=\bigl(\cos k_y,\;\mp2\tilde t\,\sin k_x\,\sin k_y,\;\cos k_x\bigr),
\label{eq:Hpm}
\end{split}
\end{equation}
where $\tilde{t} = t'/t$.
Each step implements an SU(2) rotation at each point in the RBZ:
\begin{align}
U_{-1/2}(\mathbf k)&=\exp\bigl[i\,\theta_1(\mathbf k)\,\hat{\mathbf n}_{-}(\mathbf k)\cdot\boldsymbol\sigma\bigr],
&\theta_1&=2t\,T_1\,| \mathbf h_{\mathbf{k}}|,\\
U_{+1/2}(\mathbf k)&=\exp\bigl[i\,\theta_2(\mathbf k)\,\hat{\mathbf n}_{+}(\mathbf k)\cdot\boldsymbol\sigma\bigr],
&\theta_2&=2t\,T_2\,| \mathbf h_{\mathbf{k}}|,
\end{align}
where
\begin{equation}
\hat{\mathbf n}_{\pm}(\mathbf k)=\frac{\mathbf h_{\pm1/2}(\mathbf k)}{| \mathbf h_{\mathbf{k}}|},
\end{equation}
and
\begin{equation}
| \mathbf h_{\mathbf{k}}| = \sqrt{\cos^2 k_y + 4\tilde{t}^2 \sin^2 k_x \sin^2 k_y + \cos^2 k_x}.
\end{equation}
Thus, the one-period Bloch-Floquet operator can be represented as a single equivalent SU(2) rotation
\begin{equation}
\label{eq:pipi_ideal}
U(\mathbf k, T)=U_{+1/2}(\mathbf k)\,U_{-1/2}(\mathbf k)
=\exp\bigl[i\,\Theta(\mathbf k)\,\hat{\mathbf n}_F(\mathbf k)\cdot\boldsymbol\sigma\bigr],
\end{equation}
with
\begin{equation}
\cos\Theta(\mathbf k)
=\cos\theta_1\,\cos\theta_2
  -\bigl(\hat{\mathbf n}_{+}\!\cdot\!\hat{\mathbf n}_{-}\bigr)\,
   \sin\theta_1\,\sin\theta_2,
\label{eq:cosTheta}
\end{equation}
\begin{equation}
\begin{split}
\hat{\mathbf n}_F(\mathbf k)\,\sin\Theta(\mathbf k) ={}&
  \hat{\mathbf n}_{-}\,\sin\theta_1\,\cos\theta_2
+ \hat{\mathbf n}_{+}\,\sin\theta_2\,\cos\theta_1 \\[4pt]
&\quad
+ \bigl(\hat{\mathbf n}_{-}\times\hat{\mathbf n}_{+}\bigr)\,\sin\theta_1\,\sin\theta_2.
\end{split}
\label{eq:nF}
\end{equation}
The quasienergies are then
\begin{equation}
\label{eq:pipi_quasi}
\varepsilon_{\pm}(\mathbf k)
=\pm\,\frac{\Theta(\mathbf k)}{T} + \frac{2\pi\,m}{T},
\quad m\in\mathbb Z,
\end{equation}
where \(T=T_1+T_2\) and \(\Theta(\mathbf k)\) is determined from the principal value of Eq.~\eqref{eq:cosTheta}.  The inverse cosine operation is defined modulo \(2\pi\); hence, the integer \(m\in\mathbb Z\) in Eq. \eqref{eq:pipi_quasi} accounts for the infinitely many equivalent branches in the quasienergy spectrum--an expected feature in any Floquet system.  We plot the quasienergy spectrum for the $\pm 1/2$ driving protocol on a cylindrical strip geometry for several values of $T_1$ and $T_2 = 0.2$ along with the Chern and RLBL winding numbers $W_0$ and $W_\pi$ (shorthand for $W_{\pi/T}$) in Fig. \ref{fig:Chern_pi_pi}.  We explicitly demonstrate the robustness of the flipped topological phase to finite flux ramps in Appendix \ref{app: ramps}.

\begin{figure}[t]
\label{fig:Ribbon_pi_pi}
  \centering
  \paneltagH{0.19\textheight}{a}%
  \includegraphics[width=0.4\textwidth,height=0.22\textheight,keepaspectratio]{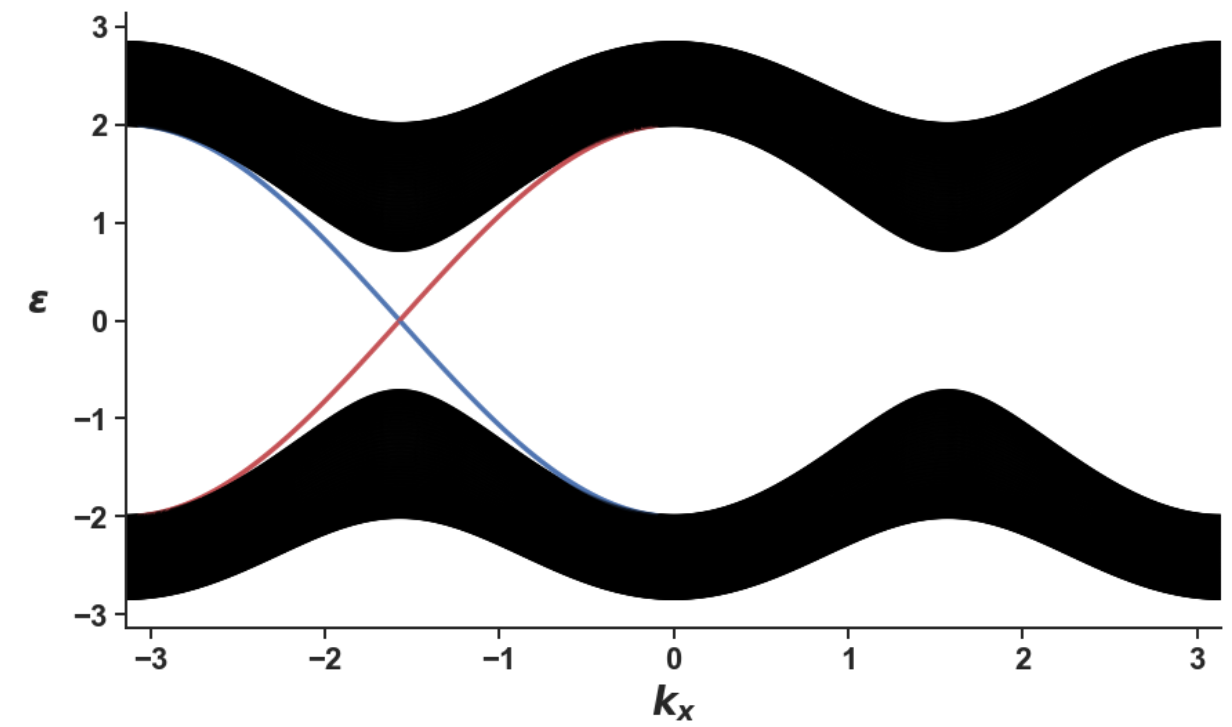}%
  \hfill
  \paneltagH{0.19\textheight}{b}%
  \includegraphics[width=0.4\textwidth,height=0.22\textheight,keepaspectratio]{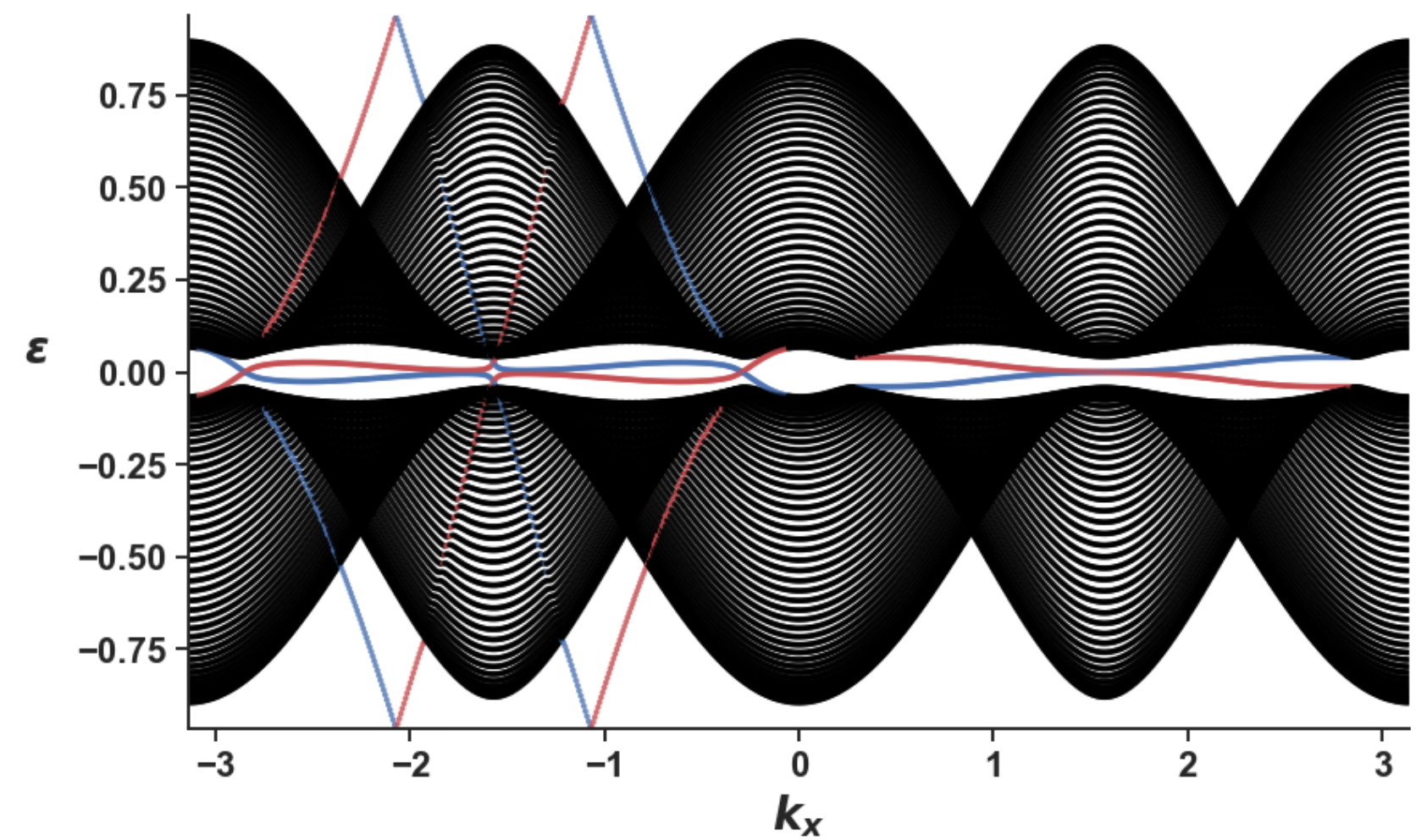}

  \caption{\raggedright
  Floquet strip spectra plotted in the principal Floquet zone, $\varepsilon \in (-\pi/T, \pi/T]$, for a cylinder of height $N_y=90$ under the $\pm1/2$ driving protocol and the associated topological phase diagrams. In the quasienergy plots the vertical axes are listed in units of $t$, the horizontal axes are in units of inverse lattice spacing, $a^{-1}$, and the red (blue) curves denote states that are localized to the top (bottom) of the cylinder. (a) $T_2=0.8$, $1/2$ flux dominates and the Chern numbers of the bottom and top bands mimic the static $\pi$-flux Harper model $(C_b, C_t)=(1,-1)$. (b) $T_2=3.05$, a band touching event leads to RLBL invariants $W_{\pi}=2$, $W_{0}=1$ and hence a Chern inversion $(C_b,C_t)=(-1,1)$.}
  \label{fig:Ribbon_pi_pi}
\end{figure}

Given that there are two quasienergy bands in the principal zone for the $-1/2 \rightarrow 1/2$ drive it is possible to derive an expression for the Chern numbers analytically. 
One finds after an explicit evaluation (see Appendix~\ref{app:chern-derivation}) that the Chern number of the ``bottom'' band in the principal Floquet zone reduces to the 
simple form when the gaps at 0 and $\pi/T$ are open 
\begin{equation}
C_b=\operatorname{sgn}\!\bigl[\sin(4 t'(T_2 - T_1))\bigr].
\label{eq:Chern_sign_rule}
\end{equation} 
Equation~\eqref{eq:Chern_sign_rule} agrees with the results plotted in Fig. \ref{fig:Ribbon_pi_pi} by employing Eq. \eqref{eq:WminusWisC}.  For $T_1 = 0.2$ and $1.4 \lessapprox T_2 < 2$ $W_0 - W_\pi = C_b = 1$, whereas for the anomalous window where $T_2 \sim 3$ we have that $W_0 - W_\pi = C_b = -1$.  When $T_1 = T_2$, or when $t' = 0$, the drive is invariant under time reversal symmetry and the bands never separate from one another.  Furthermore, from Eq. \eqref{eq:cosTheta}, the gap at 0 is closed when $\cos{\Theta(\mathbf{k})} = +1$ for at least one point in the RBZ, and the gap at $\pi/T$ is closed when $\cos{\Theta(\mathbf{k})} = -1$ for at least one point in the RBZ. Thus, the Chern numbers of both bands are well defined iff $\cos{\Theta(\mathbf{k})} \neq \pm1$. The full Chern phase diagram is plotted in Fig. \ref{fig:Chern_pi_pi}.  We note that the flipped topological phases occupy larger regions in $T_1 - T_2$ space when the NNN hopping is increased relative to the NN hopping, i.e. when $\tilde{t}$ is increased; and, in fact, for $\tilde{t}  <  0.25$ the flipped Chern phase cannot be realized at all. 

\section{\label{sec:high_f}General Drives}
Now we consider the more general case of flux switching routines in which the flux switches between arbitrary $\alpha_j$ values.  For the general case the Bulk Floquet operator is the time-ordered product
\begin{equation}
\label{eq:U_piecewise_bulk}
U(\mathbf{k}, T) = \prod_{j} U_j(\mathbf{k}, T_j)
\end{equation}
with the constituent bulk unitaries determined as
\begin{equation}
U_j(\mathbf{k}, T_j) = \text{exp}\left[-i H_j(\mathbf{k) } T_j\right],
\end{equation}
and the branch-cut-specified Floquet bulk Hamiltonian defined via
\begin{equation}
U(\mathbf{k}, T) = \text{exp}[-i H_F^{(\varepsilon_{\text{c}})}(\mathbf{k}) T],
\end{equation}
where we impose the branch cut at $\varepsilon_{\text{c}}$ in the same way as in Eq. \eqref{eq: branchcut_ham}.
The bulk Hamiltonian for the $j^{\text{th}}$ step is defined on the $Q$-folded RBZ by performing the appropriate Fourier transformations of Eq. \eqref{eq: H_j}

\begin{equation}
\label{eq:bulk_matrix}
\begin{aligned}
H_j(\mathbf k) = -2t\,D^j_{0}
\;-\; t\bigl(e^{-ik_y} S^\dagger + e^{ik_y} S\bigr)\\
\;-\; 2t'\!\left( e^{ik_y} S\,D^j_{1/2} + e^{-ik_y} S^\dagger D^j_{-1/2}\right),\\
D^j_{\delta}
= \operatorname{diag}\!\Bigl(\cos\!\bigl[k_x - 2\pi\alpha_j (m+\delta)\bigr]\Bigr)_{m=0}^{Q-1},\\
S_{mn}=\delta_{m,n+1\;(\mathrm{mod}\;Q)},\quad S^\dagger_{mn}=\delta_{m,n-1\;(\mathrm{mod}\;Q)}.
\end{aligned}
\end{equation}

\begin{figure}[t]
  \centering
  \includegraphics[width=.5\textwidth]{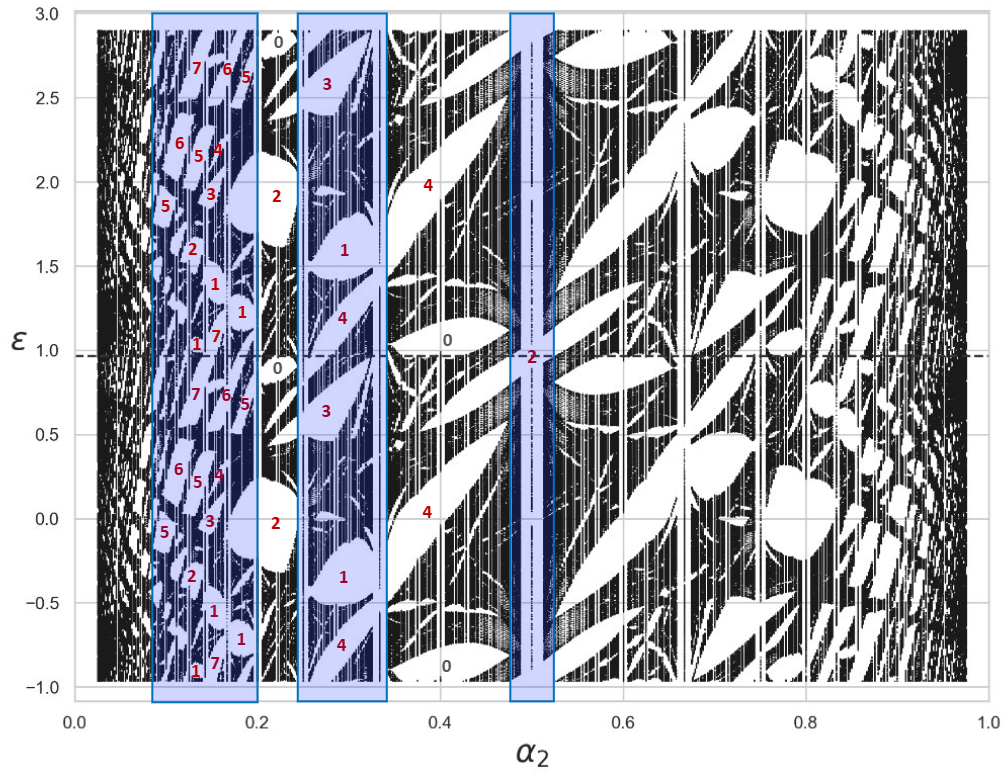}
  \caption{\raggedright
    Butterfly quasienergy spectrum for $\alpha_1 = -1/2$, $T_1 = 0.2$, $T_2 = 3.05$, $t' = 0.3$
    plotted for two adjacent Floquet zones. The dashed horizontal line indicates the top edge of the
    principal Floquet zone $\varepsilon = \pi/T$, and $\alpha_2$ values are selected from the set of
    all unique reduced fractions $0 < p_2/q_2 < 1$ with $q_{2, {\text{max}}}=40$. $W$ invariants are
    listed for a selection of the main gaps in red text.  Also highlighted
are three selected regions which feature strictly nonzero
winding in all gaps which is unrealizable in any static system. We verify agreement between the Floquet–Středa equation
    (Eq.~\eqref{eq: Streda_eq}), the approximate Suzuki–Morikawa expression (Eq.~\eqref{eq:Suzuki_W}),
    and the number of chiral edge modes in each gap.}
  \label{fig:butterfly_half}
\end{figure}

\begin{figure}[t]
  \centering
  \includegraphics[width=.5\textwidth]{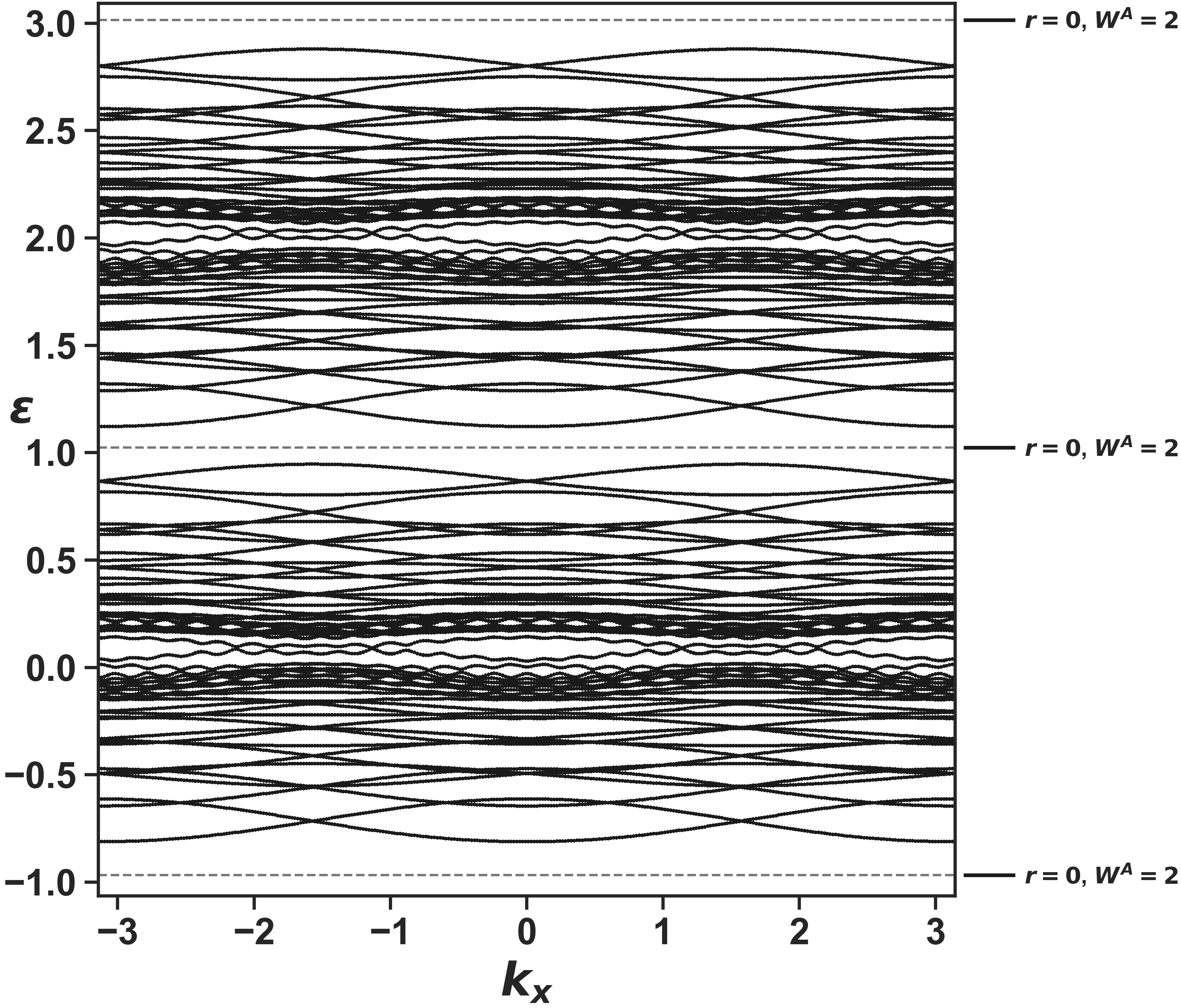}
  \caption{\raggedright
    Bulk quasienergy spectrum for the $\mathbf{T}=(0.2,\,3.05)$,\;
  $\boldsymbol{\alpha}=\!\left(-\tfrac{1}{2},\, \tfrac{12}{25}\right)$ drive in two adjacent Floquet zones and the branch cut placed at the quasienergy $\varepsilon_\text{c} = -\pi/T$.  This quasienergy spectrum demonstrates a distinctly Floquet topological phase--the total Chern number of the  bands is zero but there is an anomalous winding of $W^A = 2$ through the only distinct gap which straddles the Floquet zone boundary.}.
  \label{fig:band_labels_anom}
\end{figure}

A complete ``butterfly'' quasienergy spectrum may be obtained by diagonalizing Eq. \eqref{eq:U_piecewise_bulk} and varying the values of each $\alpha_j$.  Given $L$ distinct flux steps at fixed values of $\{T_j\}$, it is an $L$+1 dimensional object with the quasienergy gaps forming empty chambers in the space.  We plot a slice of the quasienergy butterfly spectrum for $\alpha_1 = -1/2$, $T_1 = 0.2$, and $T_2 = 3.05$ by varying $\alpha_2$ and diagonalizing $U(\mathbf{k}, T)$ in Fig. \ref{fig:butterfly_half}.  Notably, near $\alpha_2=1/2$ the gap around quasienergy $\varepsilon=\pi/T$ carries winding of $W = 2$ over a finite interval of $\alpha_2$.  Importantly, this feature is not tied to a particular
zone-edge convention: shifting the branch cut moves which gap is designated
$W_\pi$, but does not remove the underlying spectral-flow structure responsible
for $W=2$.  Moreover, the blue-highlighted regions in
Fig.~\ref{fig:butterfly_half} exhibit all gaps carrying a nonzero winding--a behavior that
cannot occur in any static band structure.  In equilibrium there is always a
topmost and bottommost gap with zero net chiral spectral flow, and no choice of
branch cut for the flux-switching protocols in the blue regions highlighted in Fig. \ref{fig:butterfly_half} can emulate trivial outer gaps.  Another noteworthy observation of Fig. \ref{fig:butterfly_half} is that if one slightly detunes $\alpha_2$ from 1/2 one yields a quasienergy spectrum with only one distinct gap (with winding $W$ = 2).  This topological phase has no static counterpart--the sum of the quasienergy bands' Chern numbers is zero, and yet there is a nonzero winding through the single distinct gap in the spectrum.  We plot the bulk quasienergy spectrum, flattened along $k_y$, for this phase in Fig. \ref{fig:band_labels_anom}.  These  observations make it clear that the flux-switching drive can therefore access topology absent in static Harper--Hofstadter physics.

In the static Harper-Hofstadter model, the St\v{r}eda formula \cite{Streda_og} may be used to relate the rate of change of occupied states with magnetic flux to the sum of the Chern numbers of the occupied bands.  Recently, the Floquet analogue of the St\v{r}eda formula was developed \cite{Streda} which we may use with particular benefit for our purposes here.  The windings through a given gap at quasienergy $\mu$ for a general driven model are partitioned into normal ($N$) and anomalous ($A$) contributions
\begin{equation}
\label{eq:Streda_eq 1}
W_\mu = W^{N,(\varepsilon_{\text{c}})}(\mu) + W^{A,(\varepsilon_{\text{c}})},
\end{equation}
where
\begin{equation}
\label{eq: Streda_eq}
W^{N,(\varepsilon_{\text{c}})}(\mu) = \frac{\partial N_{\text{eff}}^{(\varepsilon_{\text{c}})}(\mu)}{\partial \alpha} \, , \quad 
W^{A,(\varepsilon_{\text{c}})} = \frac{1}{\omega}\frac{\partial \Tr\!\left[H_{F}^{(\varepsilon_{\text{c}})}\right]}{\partial \alpha}.
\end{equation}
Here $\omega=2\pi/T$ is the frequency of the drive, $N_{\text{eff}}^{(\varepsilon_{\text{c}})}(\mu)$ counts states in the chosen Floquet zone from its ``bottom'' up to quasienergy $\mu$, and the trace is taken over the single particle Floquet Hamiltonian.  The normal contribution $W^{N,(\varepsilon_{\text{c}})}$ is a bulk invariant that quantifies the flux-induced spectral flow across a fixed quasienergy gap via the effective integrated density of states; in this sense it is the direct Floquet analogue of the static Středa response.  On the other hand, the anomalous contribution can be understood as a quantized energy flow between the system and the driving field(s) \cite{Streda}.  In the original construction of this expression $\alpha$ is treated as an artificial flux threaded through the lattice: it perturbs the base model, and the response to this flux quantifies the winding through a given gap.  Furthermore, although these equations were first implemented using the natural Floquet zone convention introduced in Ref.~\cite{Rudner2015}, Eq.s ~\eqref{eq:Streda_eq 1} and \eqref{eq: Streda_eq} apply whenever the branch cut is placed in a gap in the quasienergy spectrum \cite{Streda}, and produce results consistent with Eq.~\eqref{eq:Suzuki_W} as well as the visual count of chiral edge modes spanning each gap. Two observations make Eq.s \eqref{eq:Streda_eq 1}, \eqref{eq: Streda_eq} particularly transparent. First, $W_{\varepsilon_{\text{c}}} = W^{A,(\varepsilon_{\text{c}})}$ is the total winding through the unique gap that straddles the Floquet-zone boundary (i.e. the gap containing the branch cut).  Furthermore, the normal winding through a gap at $\mu$, $W^{N,(\varepsilon_{\text{c}})}(\mu)$, is the sum of the Chern numbers of all bands from the bottom of the chosen Floquet zone up to the quasienergy $\mu$; both of these facts follow directly from Eq. \eqref{eq:WminusWisC}, and Eq. \eqref{eq:Streda_eq 1}.

For the $L$-step arbitrary flux-switching drive Eq.~\eqref{eq: Streda_eq} may be expressed as 
\begin{equation}
\label{eq: Streda_eq_N}
W^{N,(\varepsilon_{\text{c}})}(\mu) = \mathbf{u} \cdot \boldsymbol{\nabla}_{\boldsymbol{\alpha}} N_{\text{eff}}^{(\varepsilon_{\text{c}})}(\mu)
\end{equation}
and
\begin{equation}
\label{eq: Streda_eq_A}
W^{A,(\varepsilon_{\text{c}})} = \frac{1}{\omega} \mathbf{u} \cdot \boldsymbol{\nabla}_{\boldsymbol{\alpha}} \Tr\!\left[H_{F}^{(\varepsilon_{\text{c}})}\right],
\end{equation}
where the vector $\mathbf{u}=(1,1,\ldots,1)$ has length $L$, and $\boldsymbol{\alpha}=(\alpha_1,\alpha_2,\ldots,\alpha_L)$ is the vector of dimensionless flux values attained in a period.

Let us first consider the anomalous contribution to the winding,
Eq.~\eqref{eq: Streda_eq_A}. Using the identities $\det(e^{A})=e^{\Tr A}$
and $\det(AB)=\det(A)\det(B)$ together with the branch-cut-specified version of Eq. \eqref{eq:U_piecewise}, we obtain
\begin{equation}
\label{eq:det_short}
\det U(T)=\exp\!\Bigl(-iT\,\Tr\!\bigl[H_F^{(\varepsilon_{\mathrm c})}\bigr]\Bigr)
=\exp\!\Bigl(-i\sum_{j=1}^{L}T_j\,\Tr[H_j]\Bigr).
\end{equation}
Equality of exponentials fixes $T\,\Tr[H_F^{(\varepsilon_{\mathrm c})}]$ only modulo $2\pi$.
The branch-cut logarithm selects the unique representative
\begin{equation}
\label{eq:Tr_connection_short}
T\,\Tr\!\bigl[H_F^{(\varepsilon_{\mathrm c})}\bigr]\in
\bigl(\varepsilon_{\mathrm c}T,\;\varepsilon_{\mathrm c}T+2\pi\bigr],
\end{equation}
which implies
\begin{equation}
\label{eq:Tr_connection}
\Tr\!\bigl[H_F^{(\varepsilon_{\mathrm c})}\bigr]
=\sum_{j=1}^{L}\frac{T_j}{T}\,\Tr[H_j]
+m^{(\varepsilon_{\mathrm c})}(\boldsymbol{\alpha})\,\omega,
\end{equation}
where $m^{(\varepsilon_{\mathrm c})}(\boldsymbol{\alpha}) \in \mathbb{Z}$ is the unique integer that enforces
Eq.~\eqref{eq:Tr_connection_short}.  Putting together Eqs.~\eqref{eq: Streda_eq_A} and \eqref{eq:Tr_connection} we obtain
\begin{equation}
\label{eq:WA_fluxes}
W^{A,(\varepsilon_{\text{c}})}
= \sum_{j=1}^L \frac{T_j}{2\pi}\,\partial_{\alpha_j}\Tr\!\left[H_j\right]
+ \mathbf{u}\cdot \boldsymbol{\nabla}_{\boldsymbol{\alpha}}m^{(\varepsilon_{\text{c}})}(\boldsymbol{\alpha}).
\end{equation}
For the Harper--Hofstadter model in question, the trace of each constituent
Hamiltonian vanishes (with or without ramping), i.e.\ $\Tr[H_j]=0$ for all
$\alpha_j$.  In this case Eq.~\eqref{eq:WA_fluxes} reduces to
\begin{equation}
\label{eq:WA_fluxes_final}
W^{A,(\varepsilon_{\text{c}})} = \mathbf{u}\cdot \boldsymbol{\nabla}_{\boldsymbol{\alpha}}m^{(\varepsilon_{\text{c}})}(\boldsymbol{\alpha}).
\end{equation}

Equation~\eqref{eq:WA_fluxes_final} then admits a simple bulk interpretation.
For a fixed branch cut at $\varepsilon_{\mathrm c}$, the branch-cut logarithm
selects a unique Floquet Hamiltonian
$H_F^{(\varepsilon_{\mathrm c})}(\boldsymbol{\alpha})=\frac{i}{T}\log_{-\varepsilon_{\mathrm c}T}U(T;\boldsymbol{\alpha})$.
As $\boldsymbol{\alpha}$ is varied, maintaining the same branch choice can force
some quasienergy eigenvalues to be shifted by $\pm\omega$ so that they remain in
the interval fixed by the logarithm. Such refoldings shift
$\Tr\!\left[H_F^{(\varepsilon_{\mathrm c})}\right]$ only by integer multiples of
$\omega$, which is precisely the origin of the integer
$m^{(\varepsilon_{\mathrm c})}(\boldsymbol{\alpha})$ in
Eq.~\eqref{eq:Tr_connection}. Consequently,
$W^{A,(\varepsilon_{\mathrm c})}$ counts the net number of $\pm\omega$
refoldings required as the flux protocol is changed, i.e.\ the cumulative
change of the branch index $m^{(\varepsilon_{\mathrm c})}$ under flux variation.
On a cylindrical lattice this same integer is visible as spectral flow of chiral
edge modes through the gap containing the branch cut, in direct analogy with
Ref.~\cite{HofstadterSpectralFlow}.

As for the normal contribution to the total winding, Eq.~\eqref{eq: Streda_eq_N} may be integrated to yield the Diophantine equation
\begin{equation}
N_{\text{eff}}^{(\varepsilon_{\text{c}})}(\mu) = s^{(\varepsilon_{\text{c}})}(\mu) + \sum_{j=1}^L W^{N,(\varepsilon_{\text{c}})}_j(\mu)\, \alpha_j,
\end{equation}
where $W^{N,(\varepsilon_{\text{c}})}_j(\mu)=\partial_{\alpha_j}N_{\text{eff}}^{(\varepsilon_{\text{c}})}(\mu)$ is the integer contribution of the $j^{\text{th}}$ step in the drive to the total normal winding of the gap for the chosen branch cut. Defining the effective number of states per reduced magnetic unit cell below $\mu$ in a fixed Floquet zone by $N_{\text{eff}}^{(\varepsilon_{\text{c}})}(\mu)=[r\,(\text{mod}\,Q)]/Q$ we yield the Wannier-like equation \cite{Wannier}
\begin{equation}
r\,(\text{mod}\,Q)=Q\,s^{(\varepsilon_\text{c})}(\mu)+\sum_{j = 1}^L  W^{N,(\varepsilon_{\text{c}})}_j(\mu)\,\tilde{p}_j,
\end{equation}
or, equivalently,
\begin{equation}
\label{eq:Diophantine}
r\,(\text{mod}\,Q)=\sum_{j = 1}^L  W^{N,(\varepsilon_{\text{c}})}_j(\mu)\,\tilde{p}_j\;(\text{mod}\,Q),
\end{equation}
where $\tilde{p}_j=Q\,p_j/q_j$.  In the above expressions $r$ is taken mod $Q$ to ensure applicability in the extended Floquet zone scheme, but when working in a fixed Floquet zone one may discard the mod $Q$ on the left hand side of the equation and attribute to each gap a definite $r$ value with the understanding that the top and bottom gaps are labeled the same, $r=Q$ (mod $Q$) $=0$.  We emphasize that Eq.~\eqref{eq:Diophantine} holds for any choice of branch cut $\varepsilon_{\text{c}}$ placed in a gap: changing $\varepsilon_{\text{c}}$ changes the Floquet-zone assignment and hence the integrated density of states up to $\mu$, $N_{\text{eff}}^{(\varepsilon_{\text{c}})}(\mu)$, in that zone.  This correspondingly shifts the integers $W^{N,(\varepsilon_{\text{c}})}_j(\mu)$ and $r$, while leaving the total winding $W_\mu$ invariant via the corresponding change in $W^{A,(\varepsilon_\text{c})}$. 

\begin{figure}[t]
  \centering
  \includegraphics[width=.5\textwidth]{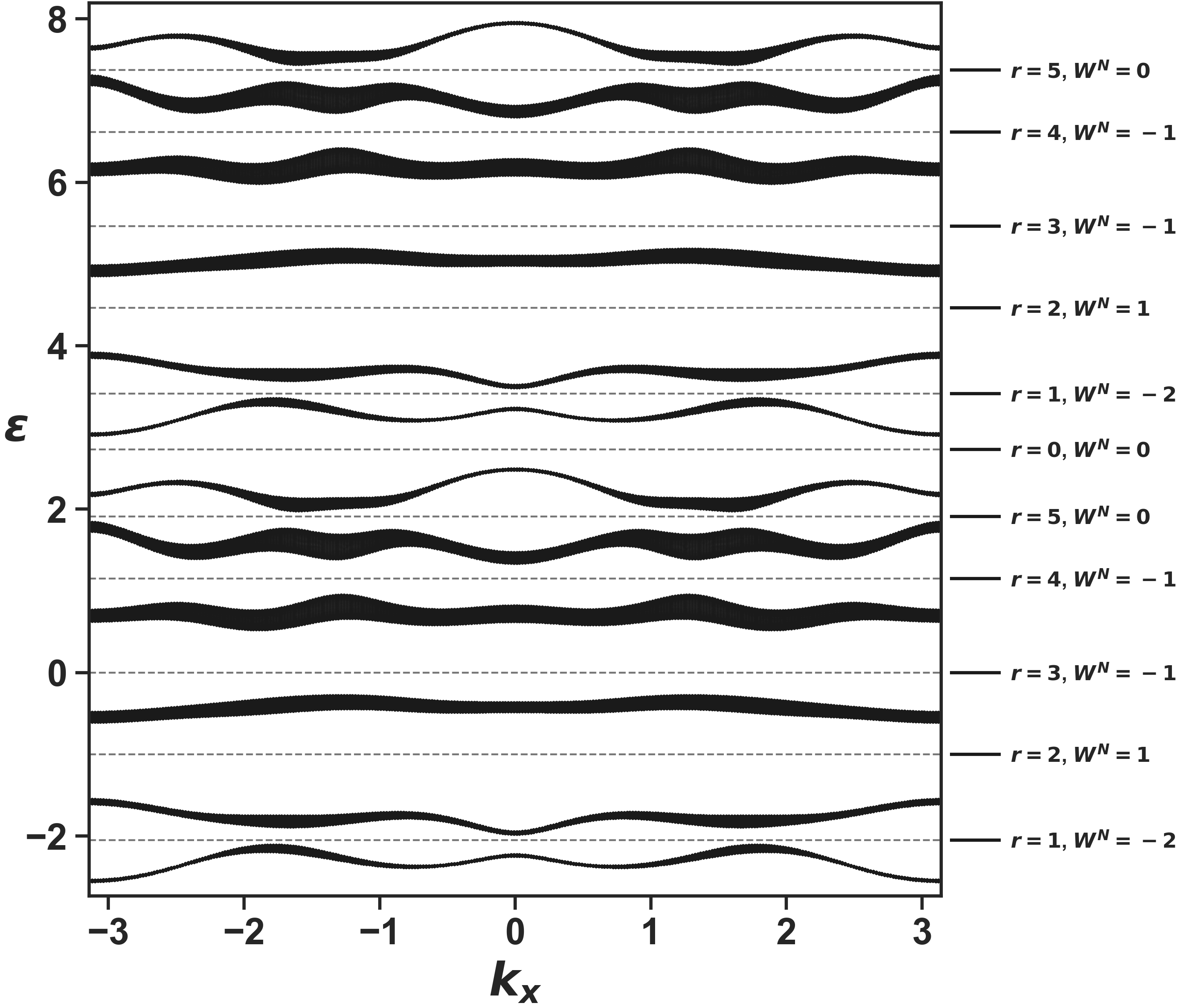}
  \caption{\raggedright
    Bulk quasienergy spectrum for the $\mathbf{T}=(0.5,\,0.5,\,0.15)$,\;
  $\boldsymbol{\alpha}=\!\left(-\tfrac{2}{3},\, -\tfrac{1}{2},\, \tfrac{1}{3}\right)$ drive with gap labels in two adjacent Floquet zones with the branch cut placed at the quasienergy $\varepsilon_\text{c} = -\pi/T$.  Within each gap the per-step windings and the corresponding Diophantine equation are explicitly listed in Table \ref{tableII}}.
  \label{fig:band_labels}
\end{figure}

We remark that this Diophantine equation is very similar in form to that which one would obtain when investigating higher-dimensional Harper-Hofstadter models \cite{MontambauxKohmoto1990,KohmotoHalperinWu1992}, but the integers here are related to a temporal contribution to a single net winding in a gap, whereas in higher dimensions the integer windings are interpreted as a Hall response associated with each distinct plane that comprises the lattice. We thus see that the gaps in the quasienergy spectrum may be labeled according to the per-step normal windings, $W_j^{N,(\varepsilon_{\text{c}})}(\mu)$ accrued in each cycle for a fixed branch cut choice, whereas the anomalous contribution to the net winding instead records the total winding through the gap which straddles the branch-cut.  We list solutions to Eq.~\eqref{eq:Diophantine} for a three-flux drive for two different sets of dwell times $\mathbf{T} = (T_1, T_2, T_3)$ in the principal Floquet zone ($\varepsilon_{\text{c}} = -\pi/T$) in
Tables~\ref{tableI} and \ref{tableII}.  We plot the bulk band dispersion of the situation illustrated in Table \ref{tableII}, flattened along $k_y$, and include gap labeling in Fig.~\ref{fig:band_labels}.

Interestingly, Eq. \eqref{eq:Diophantine} applies to any distinct flux switching routine regardless of the dwell times involved.   This is illustrated in Tables~\ref{tableI} and \ref{tableII} where the winding in the first gap is different ($W^N = -2$ vs. $W^N = 0$) due to the different dwell times, but the underlying Diophantine equation used to label the gaps is still obeyed.

\noindent
\begin{minipage}[t]{0.48\textwidth}
  \refstepcounter{table}\label{tableI}
  {Table \thetable.} 

  \vspace{0.6ex}
  \centering
  \setlength{\tabcolsep}{8pt}
  \begin{tabular}{c c c c c c}
  \hline
  $W^N$ & $W_1^{N}$ & $W_2^{N}$ & $W_3^{N}$ & $\Sigma$ & $(\Sigma \bmod Q)=r$ \\
  \hline
   0 & \phantom{-}0 & \phantom{-}1 & \phantom{-}-1 & \phantom{-}-5 & 1 \\
   1 & \phantom{-}0 & 0             & 1             & 2             & 2 \\
  -1 & 0             & \phantom{-}-1 & 0             & \phantom{-}3  & 3 \\
  -1 & \phantom{-}0 & 0             & -1            & \phantom{-}-2 & 4 \\
   0 & 0             & \phantom{-}-1 & 1             & 5             & 5 \\
  \hline
  \end{tabular}

  \vspace{0.6ex}
  {\footnotesize
  $\mathbf{T}=(0.2,\,0.6,\,0.4)$,\;
  $\boldsymbol{\alpha}=\!\left(-\tfrac{2}{3},\, -\tfrac{1}{2},\, \tfrac{1}{3}\right)$.\;
  $Q=\mathrm{lcm}(3,2,3)=6$;\;
  $(\tilde p_1,\tilde p_2,\tilde p_3)=(-4,-3,2)$.\;
  $r = \sum_j W_j^{N}\tilde p_j \pmod{Q}$;\;
  $\Sigma=-4W_1^{N}-3W_2^{N}+2W_3^{N}$.}
\end{minipage}
\hfill
\begin{minipage}[t]{0.48\textwidth}
  \refstepcounter{table}\label{tableII}
  {Table \thetable.} 

  \vspace{0.6ex}
  \centering
  \setlength{\tabcolsep}{8pt}
  \begin{tabular}{c c c c c c}
  \hline
  $W^N$ & $W_1^{N}$ & $W_2^{N}$ & $W_3^{N}$ & $\Sigma$ & $(\Sigma \bmod Q)=r$ \\
  \hline
  -2 & \phantom{-}-1 & \phantom{-}-1 & \phantom{-}0 & \phantom{-}7  & 1 \\
   1 & \phantom{-}1  & 0             & 0            & -4            & 2 \\
  -1 & 0             & \phantom{-}-1 & 0            & \phantom{-}3  & 3 \\
  -1 & \phantom{-}-1 & 0             & 0            & \phantom{-}4  & 4 \\
   0 & 1             & \phantom{-}-1 & 0            & -1            & 5 \\
  \hline
  \end{tabular}

  \vspace{0.6ex}
  {\footnotesize
  $\mathbf{T}=(0.5,\,0.5,\,0.15)$,\;
  $\boldsymbol{\alpha}=\!\left(-\tfrac{2}{3},\, -\tfrac{1}{2},\, \tfrac{1}{3}\right)$.\;
  $Q=\mathrm{lcm}(3,2,3)=6$;\;
  $(\tilde p_1,\tilde p_2,\tilde p_3)=(-4,-3,2)$.\;
  $r \equiv \sum_j W_j^{N}\tilde p_j \pmod{Q}$;\;
  $\Sigma=-4W_1^{N}-3W_2^{N}+2W_3^{N}$.}
\end{minipage}

\section{Discussion}
In this work we studied the topology and gap organization of Floquet flux–switching drives via the Harper–Hofstadter model. Generically, for a periodic drive comprised of a series of fluxes $\{p_j/q_j\}$, the quasienergies are organized into $Q=\mathrm{lcm}\{q_j\}$ magnetic subbands.  For the simplest nontrivial case of the two-step $-1/2 \rightarrow 1/2$ flux drive with nonzero NNN hopping the bulk Floquet operator, $U(\mathbf{k}, T)$, can be interpreted as a single net SU(2) rotation comprised of two distinct rotations with individual rotation angles proportional to the dwell times spent at -1/2, and +1/2 flux.  We determined the Chern numbers for the upper and lower bands to be
\[
C=\mp \operatorname{sgn}\!\left[\sin\!\left(4 t'(T_2-T_1)\right)\right],
\]
valid whenever the $0$ and $\pi/T$ gaps are open, and where the upper and lower signs correspond to the top and bottom bands in the principal Floquet zone respectively. This formula captures the Chern inversion observed in our strip spectra and agrees with the numerically calculated RLBL windings in the regimes where $W_\pi = 2$.   

The bulk Floquet operator for a general flux switching drive may be interpreted as a series of U($Q$) rotations, where the culmination of all terms in a product gives one single equivalent rotation, $U(\mathbf{k}, T)$.  The quasienergies for the general case were found to have a Wannier-like labeling structure via the adapted Floquet–St\v{r}eda formula, Eq. \eqref{eq: Streda_eq}.  Namely, we found a Diophantine equation that relates gap labels in the quasienergy spectrum to the per-step ``normal'' windings explored during the periodic flux drive.  The winding in any gap is partitioned into normal and anomalous contributions via $W=W^{\!N}+W^{\!A}$; the normal contribution and the fluxes of the drives are combined into a $L$-dimensional Diophantine equation for an $L$-step drive as
\begin{equation}
r\,(\text{mod}\,Q)=\sum_{j = 1}^L  W^{N,(\varepsilon_{\text{c}})}_j(\mu)\,\tilde{p}_j\;(\text{mod}\,Q),
\end{equation}
where the total normal winding through the gap at quasienergy $\mu$ with the branch cut at $\varepsilon_\text{c}$ is the sum of the per-step windings $W^{N, (\varepsilon_{\text{c}})}(\mu) = \sum_{j=1}^L W^{N, (\varepsilon_{\text{c}})}_j(\mu)$, $\tilde{p}_j = Q p_j/q_j$, and the mod $Q$ on the left hand side of the equality ensures the relation holds in the extended Floquet zone.  Given this formulation, it is understood that the top and bottom gaps share $r = 0$.  Notably, this expression is entirely agnostic to the frequency of the actual drive--it defines a congruence that must be followed for a given set of flux quenches, and thus it may be understood as a rule which dictates the allowed transfer of ``normal'' topological charge (not accounting for a changing $W^{A,(\varepsilon_{\text{c}})}$ which may occur only when the gap straddling the branch cut closes and reopens) across band touching events as dwell times are tuned.  Hence, the set of windings featured in the spectrum may be engineered in a predictable fashion by selecting a set of fluxes in a drive and tuning the respective dwell times.   Because $W^{A,(\varepsilon_{\text{c}})}$ is not featured in this expression the ``rule-set'' that dictates how $W^{A,(\varepsilon_{\text{c}})}$ can change between driving periods is different--a subject for future investigation.

An important practical question is how the per-step normal windings
$W^{N,(\varepsilon_{\mathrm c})}_j(\mu)$ entering the Diophantine labeling might be accessed
experimentally.  Equation~\eqref{eq: Streda_eq_N} provides an operational route:
$W^{N,(\varepsilon_{\mathrm c})}_j(\mu)$ is the response of the effective integrated density of states
$N_{\mathrm{eff}}^{(\varepsilon_{\mathrm c})}(\mu)$ to the flux parameter in the $j^{\text{th}}$ segment of the drive,
with all other steps held fixed.  Formally, one may therefore compare two nearby flux-switching
protocols that differ only by $\alpha_j\!\to\!\alpha_j+\delta\alpha$ (at fixed dwell times and fixed branch cut),
and extract $W^{N,(\varepsilon_{\mathrm c})}_j(\mu)$ from a finite difference of
$N_{\mathrm{eff}}^{(\varepsilon_{\mathrm c})}(\mu)$.  In practice, the key experimental ingredient is access to a
well-defined occupation function for Floquet states relative to the chosen branch cut---a nontrivial requirement
for an isolated driven system.  A concrete route is to infer $N_{\mathrm{eff}}$ from particle-density measurements
in a steady state stabilized by coupling to reservoirs or engineered dissipation, as advocated in recent proposals
for Floquet--St\v{r}eda responses~\cite{Streda}.  Cold-atom platforms already provide much of the needed toolbox:
topological response has been measured in cold-atom Hofstadter bands~\cite{Aidelsburger2015Chern}, and density-based St\v{r}eda-type
detection schemes have been proposed using trapped gases via bulk density profiles~\cite{Umucalilar2008StredaDensity}.
Moreover, cold-atom two-terminal ``atomtronic'' geometries realize systems coupled to macroscopic reservoirs with
controlled chemical potentials and in situ density readout~\cite{Brantut2012Conduction}, providing a natural setting
for the reservoir-stabilized measurements envisioned in Ref.~\cite{Streda}.  Complementary information about gap
structure and spectral flow can be obtained from real-space imaging of edge/bulk structure in synthetic Hall ribbons~\cite{Mancini2015ChiralRibbons,Stuhl2015EdgeStates}
and from quantum-gas microscopy in Harper--Hofstadter settings~\cite{Tai2017MicroscopyHH}.  Repeating the finite-difference
procedure step-by-step, varying only a single $\alpha_j$ in the protocol at a time, would then reconstruct the vector of per-step responses and
therefore the per-step windings that label gaps in multi-step flux-switching quasienergy spectra.  Beyond the per-step normal responses, the full gap winding $W_{\varepsilon}$ (and hence the anomalous contribution $W_A$) can be accessed on cold-atom platforms by combining quasienergy-gap measurements with local Hall-deflection measurements, as demonstrated experimentally in an anomalous Floquet optical-lattice realization where the complete set of Floquet invariants was inferred \cite{Wintersperger2020AnomalousFloquet}.  Alternatively, a direct measurement protocol for the Floquet gap winding numbers in driven optical lattices---based on tracking and tomographically identifying band-touching singularities along a one-parameter family of experimentally feasible drives---can be applied by choosing the present flux-switching protocol as the target endpoint of that deformation \cite{Unal2019DirectMeasureFloquet}.

Future investigations should also consider time-reversal symmetric drives with spinful particles of the form $U(T) = U_{-\alpha_1}(T_1) U_{-\alpha_2}(T_2) ... U_{\alpha_2}(T_2) U_{\alpha_1}(T_1)$ as these protocols are candidates for hosting nontrivial $\mathbb{Z}_2$ invariants which protect helical edge states.  Furthermore, although nonmagnetic weak disorder is not expected to destroy the topology of the bands in this drive (as it does not destroy the topology of the static Harper-Hofstadter bands), the effects of strong disorder and particle-particle interactions in the Bose-Hubbard or Fermi-Hubbard models are certainly interesting future directions to explore.

\section*{Acknowledgments}
This research was generously supported by the William and Linda Frost Fund
in the Cal Poly Bailey College of Science and Mathematics.

\section*{Data availability}
The data are not publicly available. The data are
available from the authors upon reasonable request.

\appendix

\section{Ramping terms in the $-1/2\!\to\!1/2$ flux-switching routine}
\label{app: ramps}
\begin{figure}[]
    \centering
    \includegraphics[width=\columnwidth]{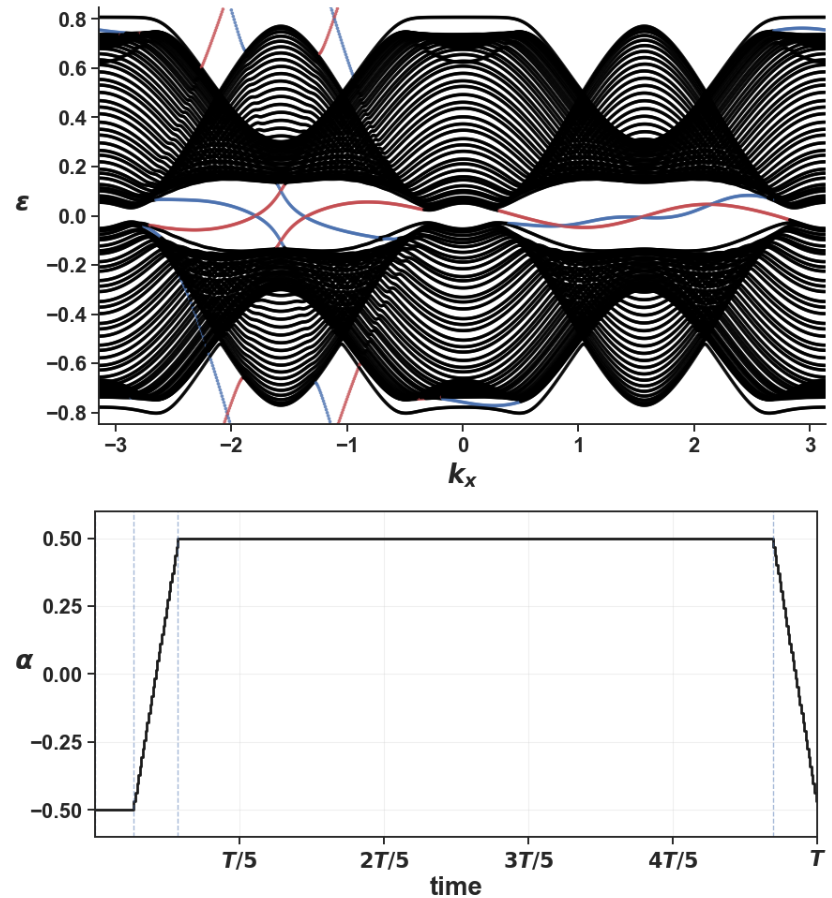}
    \caption{\raggedright
    $-1/2 \rightarrow 1/2$ flux-switching quasienergy spectrum and ramp schedule with $T_1=0.2$, $T_2=3.05$, $t'=0.3$, $t=1$, and $\tau=0.225$. Thirty terms at equally spaced flux values between $-1/2$ and $+1/2$ are used for each ramp. The primary gaps close when ramp times exceed approximately $\tau\approx 0.35$.}
    \label{fig:ramp_fig}
\end{figure}

To investigate the effect of explicitly including ramping terms in the $-1/2 \rightarrow 1/2$ switching routine, we modify the ideal Floquet operator in Eq.~\eqref{eq:pipi_ideal} to be
\begin{equation}
U(\mathbf{k}, T)
= U_{\text{d}}(\mathbf{k}, \tau)\, U_{+1/2}(\mathbf{k}, T_2)\, U_{\text{u}}(\mathbf{k}, \tau)\, U_{-1/2}(\mathbf{k}, T_1),
\end{equation}
where the up- and down-ramp evolution operators are given as the time-ordered products
\begin{equation}
U_{\text{u}}(\mathbf{k}, \tau) = \prod_{j=1}^{N-1} U_j(\mathbf{k}, \delta\tau), \qquad
U_{\text{d}}(\mathbf{k}, \tau) = \prod_{j=1}^{N-1} U_{N-j}(\mathbf{k}, \delta\tau),
\end{equation}
with constituent unitaries
\begin{equation}
U_j(\mathbf{k}, \delta\tau) = \exp\!\left[-i\, H_{-1/2 + j/N}(\mathbf{k})\,(\tau/N)\right],
\end{equation}
and the corresponding Hamiltonians computed via Eq.~\eqref{eq:bulk_matrix} using the appropriate value of $Q$ given the number of steps in the ramp. We plot the ramp-modified quasienergy spectrum for $\tau=0.225$ in Fig.~\ref{fig:ramp_fig}. Notably, the two primary quasienergy gaps survive the inclusion of the ramping unitaries.

\section{Dependence of $W_\varepsilon$ on grid resolution when implementing Eq. \eqref{eq:Suzuki_W}}
\label{app:convergence}
\begin{figure}[]
    \centering
    \includegraphics[width=0.95\linewidth]{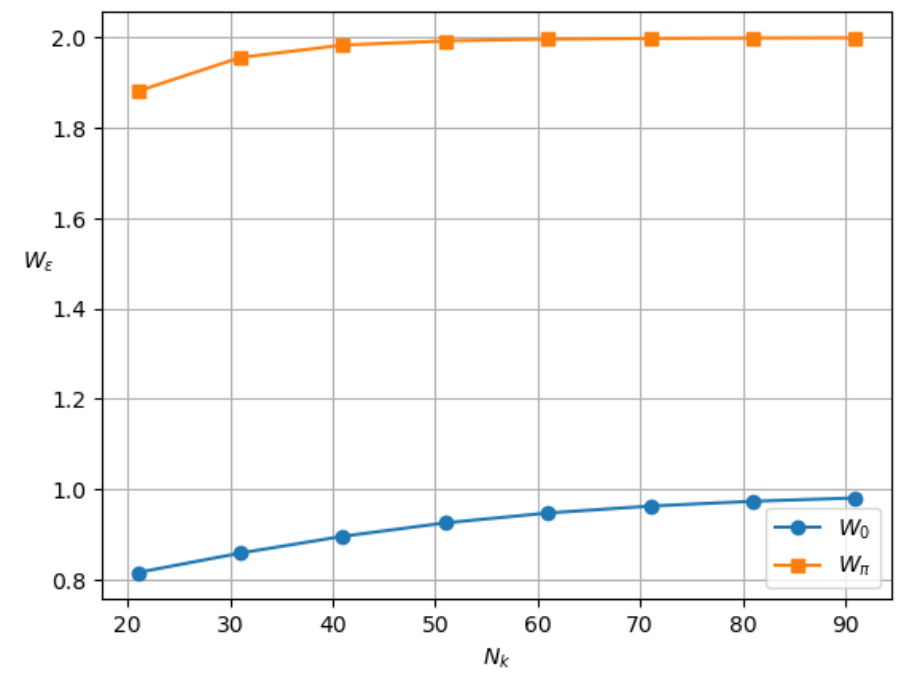}
    \caption{Calculated values of $W_0$ and $W_\pi$ for the $-\tfrac{1}{2}\!\rightarrow\!+\tfrac{1}{2}$ flux-switching protocol in the flipped Chern regime computed using $N_{k_x} \times N_{k_y} \times N_{t} = N_{k}^2 \times N_{t}$ grid resolutions using Eq. \eqref{eq:Suzuki_W}. Here $N_t = \text{round}\left[\left(\frac{51}{91}\right) N_k\right]$}
    \label{fig: conv}
\end{figure}

\begin{figure}
    \centering
    \includegraphics[width=0.95\linewidth]{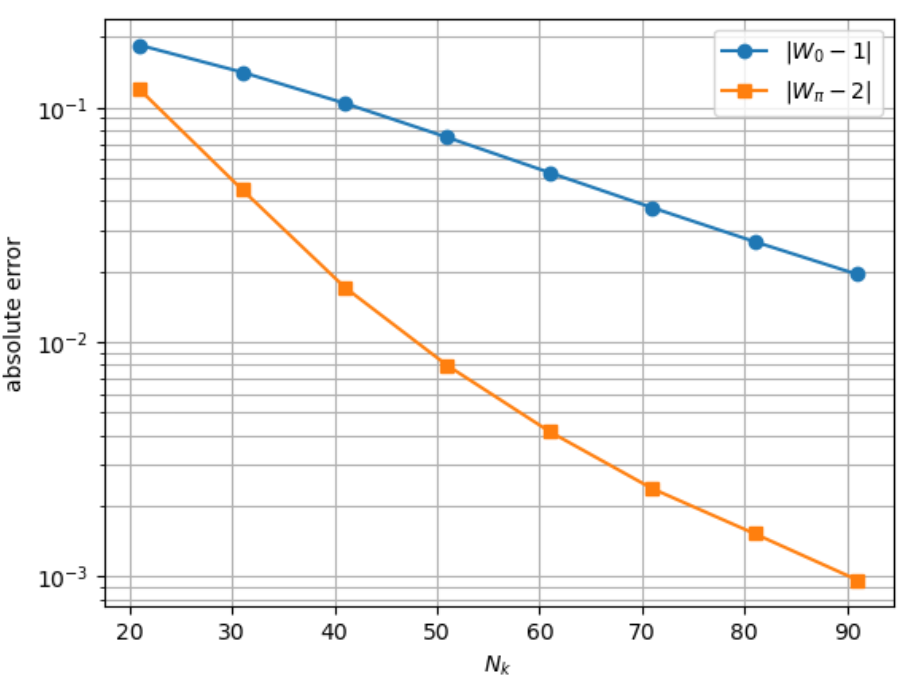}
    \caption{Calculated values of error in $W_0$ and $W_\pi$ for the $-\tfrac{1}{2}\!\rightarrow\!+\tfrac{1}{2}$ flux-switching protocol in the flipped Chern regime computed using $N_{k_x} \times N_{k_y} \times N_{t} = N_{k}^2 \times N_{t}$ grid resolutions using Eq. \eqref{eq:Suzuki_W}. Here $N_t = \text{round}\left[\left(\frac{51}{91}\right) N_k\right]$}
    \label{fig: conv2}
\end{figure}

We compute the Rudner--Lindner--Berg--Levin
(RLBL) winding number $W_\varepsilon$ by evaluating Eq. \eqref{eq:Suzuki_W} on a uniform cubic grid
on the three--torus $(k_x,k_y,t)\in \mathrm{BZ}\times[0,T)$ with periodic
boundary conditions in all three directions.  In generating Fig. \ref{fig:Chern_pi_pi} we used a
$91\times 91\times 51$ grid for all reported values of $W_\varepsilon$.
Because $W_\varepsilon$ is an integer--valued topological invariant when
the quasienergy gap at $\varepsilon$ remains open, numerical errors arise primarily due to  discretization, but steadily decrease as grid resolution is increased.  In Fig.s \ref{fig: conv}, \ref{fig: conv2} we plot $W_\varepsilon$ as computed using Eq. \eqref{eq:Suzuki_W} and the associated error for the $-\tfrac{1}{2}\!\rightarrow\!+\tfrac{1}{2}$ flux-switching protocol in the flipped Chern phase at dwell times $T_1 = 0.2$, $T_2 = 3.05$.  The values of $W_0$ and $W_{\pi}$ are stable across the largest grids considered, and the absolute errors
$|W_0-1|$ and $|W_{\pi}-2|$ decrease rapidly with increasing resolution.


\section{$-\tfrac{1}{2}\!\rightarrow\!+\tfrac{1}{2}$ flux-switching Chern numbers}
\label{app:chern-derivation}
For any two-level Hamiltonian \(H(\mathbf k)=\mathbf h(\mathbf k)\!\cdot\!\boldsymbol\sigma\), the Chern number of one band can be obtained from the index formula \cite{Sticlet2012PRB}:
\begin{equation}
\label{eq:sticlet}
C \;=\; \frac12 \sum_{\mathbf k_\alpha\in \mathcal D_i}
\operatorname{sgn}\!\Big[\big(\partial_{k_x}\mathbf h \times \partial_{k_y}\mathbf h\big)_i\Big]\,
\operatorname{sgn}\!\big(h_i(\mathbf k_\alpha)\big),
\end{equation}
where \(i\in\{x,y,z\}\) is arbitrary and
\(\mathcal D_i=\{\mathbf k:\ h_j(\mathbf k)=h_\ell(\mathbf k)=0\}\) with \(\{i,j,\ell\}=\{x,y,z\}\).

For the two-step protocol \( -\tfrac{1}{2}\rightarrow+\tfrac{1}{2} \) with dwell times \(T_1,T_2\), we define
\[
\mathbf h_{\pm}(\mathbf k)=\bigl(\cos k_y,\ \mp\,2\tilde t\,\sin k_x\sin k_y,\ \cos k_x\bigr),\qquad
\tilde t=\frac{t'}{t},
\]
\[
|\mathbf h|=\sqrt{\cos^2k_x+\cos^2k_y+4\tilde t^2\sin^2k_x\sin^2k_y}\,.
\]
The bulk Floquet operator is \(U(\mathbf k, T)=U_{+}(\mathbf k, T_2)\,U_{-}(\mathbf k, T_1)\), where
\begin{align*}
U(\mathbf{k}, T) &= e^{i \Theta(\mathbf{k})\hat{\mathbf{n}}_F \cdot \boldsymbol{\sigma}}, \quad
U_{-}(\mathbf k, T_1)=e^{\,i\theta_1(\mathbf k)\,\hat{\mathbf n}_{-}\!\cdot\!\boldsymbol\sigma},\\
U_{+}(\mathbf k, T_2)&=e^{\,i\theta_2(\mathbf k)\,\hat{\mathbf n}_{+}\!\cdot\!\boldsymbol\sigma}, \quad
\theta_{1,2}=2t\,T_{1,2}\,|\mathbf h|,\quad
\hat{\mathbf n}_{\pm}=\frac{\mathbf h_{\pm}}{|\mathbf h|}.
\end{align*}
From the SU(2) product rule,
\begin{align}
\hat{\mathbf n}_F\sin\Theta&=\hat{\mathbf n}_{-}\sin\theta_1\cos\theta_2 \notag\\
&\quad+\hat{\mathbf n}_{+}\sin\theta_2\cos\theta_1
+(\hat{\mathbf n}_{-}\!\times\!\hat{\mathbf n}_{+})\sin\theta_1\sin\theta_2,
\end{align}
and noting \((\hat{\mathbf n}_{-}\!\times\!\hat{\mathbf n}_{+})_y=0\) here (only the \(y\) component flips between steps), the \(y\) component reduces to
\begin{equation}
\label{eq:nFy-rel}
n_{F,y}(\mathbf k)\,\sin\Theta(\mathbf k)
=\frac{2\tilde t\,\sin k_x\sin k_y}{|\mathbf h|}\,\sin\!\big(\theta_1(\mathbf k)-\theta_2(\mathbf k)\big).
\end{equation}

Choosing \(i=y\) in Eq.~\eqref{eq:sticlet} and defining the flattened \(\mathbf h=\hat{\mathbf n}_F\), we proceed using \(\mathcal D_y=\{\mathbf k:\ n_{F,x}=n_{F,z}=0\}\):
\begin{equation}
\label{eq:Dy}
\mathcal D_y \;=\; \bigl\{\,\mathbf k:\ \cos k_x=0,\ \cos k_y=0\,\bigr\}.
\end{equation}
Within the RBZ this corresponds to the two points
\[
\mathbf k_\alpha\in\Bigl\{\bigl(+\tfrac{\pi}{2},\,\tfrac{\pi}{2}\bigr),\ \bigl(-\tfrac{\pi}{2},\,\tfrac{\pi}{2}\bigr)\Bigr\}.
\]
At these points \( |\mathbf h(\mathbf k_\alpha)|=2|\tilde t| \), hence
\[
\theta_2(\mathbf k_\alpha)-\theta_1(\mathbf k_\alpha)
=4\,\operatorname{sgn}(t)\,|t'|\,(T_2-T_1).
\]

From Eq.~\eqref{eq:nFy-rel}, choosing the branch with \(\sin\Theta>0\) gives
\begin{equation}
\label{eq:sign-hy}
\operatorname{sgn}\!\big(n_{F,y}(\mathbf k_\alpha)\big)
=\operatorname{sgn}(\tilde t)\,s_\alpha\ \operatorname{sgn}\!\big(\sin(\theta_1-\theta_2)\big),
\end{equation}
where
\[
s_\alpha=\operatorname{sgn}\!\big(\sin k_x^\alpha\sin k_y^\alpha\big)\in\{+1,-1\}.
\]
A linearization of \((n_{F,x},n_{F,z})\) around \(\mathbf k_\alpha\) gives the local orientation factor
\begin{equation}
\label{eq:sign-J}
\operatorname{sgn}\!\Big[\big(\partial_{k_x}\hat{\mathbf n}_F\times\partial_{k_y}\hat{\mathbf n}_F\big)_y\Big]_{\mathbf k_\alpha}
= \,s_\alpha,
\end{equation}
so each point contributes with the same overall sign after multiplying Eqs.~\eqref{eq:sign-hy}--\eqref{eq:sign-J}.

Using Eq.~\eqref{eq:sticlet} with the flattened \(\mathbf h=\hat{\mathbf n}_F\) and summing over the two \(\mathbf k_\alpha\), the Chern number for the lower band of $\hat{\mathbf n}_F$ is
\[
C_{\mathrm{lower}}= -\operatorname{sgn}(\tilde t)\ \operatorname{sgn}\!\big(\sin(\theta_2-\theta_1)\big).
\]
The bottom Floquet band (negative quasienergy in the principal Floquet zone) corresponds to the \(+\) eigenstate of $\hat{\mathbf n}_F$ via $U(\mathbf{k},T)=e^{-iH_FT}$, so its Chern number is
\[
C_b \;=\; -\,C_{\mathrm{lower}}
\;=\; \operatorname{sgn}(\tilde t)\ \operatorname{sgn}\!\big(\sin(\theta_2-\theta_1)\big).
\]
Evaluated at \(\mathbf k_\alpha\) this yields
\begin{equation}
\label{eq:Cb-final}
C_b= \operatorname{sgn}\!\Big(\sin\!\big[\,4\,t'\,(T_2-T_1)\big]\Big),
\end{equation}
valid whenever the gaps at \(0\) and \(\pi/T\) are open.
\newpage

\bibliographystyle{apsrev4-2}
\bibliography{biblio}

\end{document}